\documentclass[preprint,showpacs,preprintnumbers,amsmath,amssymb]{revtex4}

\usepackage{graphicx}
\usepackage{dcolumn}
\usepackage{bm}
\usepackage{color}

\usepackage{feynmf}
\usepackage{slashed}
\usepackage{enumerate}

\usepackage[utf8]{inputenc}

\begin{document}

\title{Some Phenomenologies of a Simple Scotogenic Inverse Seesaw Model}

%\author{Pyungwon Ko}
%\thanks{pko@kias.re.kr}
%\affiliation{School of Physics, KIAS, 85 Hoegiro, Seoul 02455, Republic of Korea}
%\affiliation{Quantum Universe Center, KIAS, 85 Hoegiro, Seoul 02455, Republic of Korea}

\author{Yi-Lei Tang}
\thanks{tangyilei@kias.re.kr}
\affiliation{School of Physics, KIAS, 85 Hoegiro, Seoul 02455, Republic of Korea}

\date{\today}

\begin{abstract}

In this paper, we discuss and calculate the electroweak parameters $R_l$, $A_l$, and $N_{\nu}^l$ in a model that combine inverse seesaw with the scotogenic model. Dark matter relic density is also considered. Due to the stringent constraint from the ATLAS experimental data, it is difficult to detect the loop effect on $R_l$, $A_l$ in this model considering both the theoretical and future experimental uncertainties. However, $N_{\nu}^l$ can sometimes become large enough for the future experiments to verify.

\end{abstract}


\maketitle
\section{Introduction}

The Type I Seesaw mechanisms \cite{SeeSaw1, SeeSaw2, SeeSaw3, SeeSaw4, SeeSaw5} are utilized to explain the smallness of the neutrino masses by introducing some extremely heavy right-handed neutrinos with the masses $10^{8\text{-}12} \text{ TeV}$, which is far beyond the ability of any current or proposed collider facility. Suppressing the mass scales of the right-handed neutrinos below $1 \text{ TeV}$ will also lead to tiny Yukawa couplings ($\sim 10^{(-7)\text{-}(-9)}$), making it rather difficult to produce any experimental signals in reality.

The scotogenic model \cite{Scotogenic1, ScotogenicRequired, Scotogenic2, Scotogenic3} and the inverse seesaw model \cite{Inverse1, Inverse2, Inverse3, Inverse4} are the two different approaches toward the TeV-scale phenomenology corresponding to the neutrino sector. In the various versions of the scotogenic model, the active neutrinos acquire masses through loop corrections. In this case, the loop factor naturally suppresses the Majorana masses of the left-handed neutrinos. As for the inverse seesaw model,  two groups of so-called ``pseudo-Dirac'' sterile neutrinos are introduced. The contributions from the large Yukawa couplings to the left-handed neutrino masses are nearly cancelled out, with a small remnant left over due to the small Majorana masses among the pseudo-Dirac sterile neutrinos which softly break the lepton number.

As far as we know about the literature, the combination of these two models can date back to Ref.~\cite{InverseScotogenic1}, which appeared shortly after the Ref.~\cite{Scotogenic1}. There are also various papers in the literature, suggesting different variants or discussing the phenomenologies (For some examples, see Refs.~\cite{InverseScotogenic2, InverseScotogenic3, InverseScotogenic4, InverseScotogenic5}, while Ref.~\cite{DasRequired} had discussed a similar linear seesaw model.). In this paper, we discuss about a simple version of such kind of models motivated by avoiding some tight restrictions on the Yukawa coupling orders. In the usual scotogenic models, Yukawa couplings are usually constrained by the leptonic flavour changing neutral current (FCNC) such as the $\mu \rightarrow e \gamma$ bound. In the case of the inverse seesaw mechanisms, the invisible decay width of the Z boson also exert limits on the Yukawa couplings. This lead to the mixings between the active neutrinos and the sterile neutrinos and will result in the corrections to the $Z \rightarrow \nu \nu$ branching ratios on tree-level.  Combining these two models can reach some relatively larger Yukawa couplings, while evading some constraints at the same time.

\section{Model Descriptions}

The scotogenic model is based on the inert two Higgs doublet model (ITHDM). In this model, two $SU(2)_L$ Higgs doublets $\Phi_1$ and $\Phi_2$ are introduced. Let $\Phi_2$ be $Z_2$-odd, while $\Phi_1$ together with other standard model (SM) fields be $Z_2$-even, the potential for the Higgs sector is given by
\begin{eqnarray}
  V&=& m_1^2 \Phi_1^{\dagger} \Phi_1 + m_2^2 \Phi_2^{\dagger} \Phi_2 + \frac{\lambda_1}{2} (\Phi_1^{\dagger} \Phi_1)^2 + \frac{\lambda_2}{2} (\Phi_2^{\dagger} \Phi_2)^2  +  \lambda_3 (\Phi_1^{\dagger} \Phi_1)(\Phi_2^{\dagger} \Phi_2) \nonumber \\
& + & \lambda_4 (\Phi_1^{\dagger} \Phi_2)(\Phi_2^{\dagger} \Phi_1) + \frac{\lambda_5}{2} \left[ (\Phi_1^{\dagger} \Phi_2)^2 + (\Phi_2^{\dagger} \Phi_1)^2 \right],
\end{eqnarray}
where $\Phi_{1,2}$ are the two Higgs doublets with the hypercharge $Y=\frac{1}{2}$, $\lambda_{1\text{-}5}$ are the coupling constants, $m_1^2$, $m_2^2$ are the mass parameters.

In the ITHDM, only $\Phi_1$ acquires the electroweak vacuum expectation value (VEV) $v$ and the standard model (SM) Higgs $h$ originates from this doublet. All the elements of the $\Phi_2$ form the other scalar bosons $H^{\pm}$, $H$, $A$, and no mixing between the SM Higgs and the exotic bosons takes place. Therefore,
\begin{eqnarray}
\Phi_1 =  \left(
\begin{array}{c}
G^{+} \\
\frac{v + h + i G^0}{\sqrt{2}}
\end{array}
\right),~
\Phi_2 =  \left(
\begin{array}{c}
H^+ \\
\frac{H + i A}{\sqrt{2}}
\end{array} \right).
\end{eqnarray}

Due to the $Z_2$ symmetry, all the fermions $Q_L$, $u_R$, $d_R$, $L_L$, $e_R$ only couple with the $\Phi_1$ field
\begin{eqnarray}
\mathcal{L}_{\text{Yukawa}}^{\text{SM}} = -Y_{u i j} \overline{Q}_{L i} \tilde{\Phi}_1 u_{R j} - Y_{d i j} \overline{Q}_{L i} \Phi_1 d_{R j} - Y_{l i j} \overline{L}_{L i} \Phi_1 l_{R j} + \text{h.c.},
\end{eqnarray}
where $Y_{u,d,l}$ are the $3 \times 3$ coupling constants. 

The  $Z_2$-odd pseudo-Dirac sterile neutrinos $N_i = P_L N_{L i} + P_R N_{R i}$, ($i=1\text{-}3$, $P_{L,R}=\frac{1 \mp \gamma^5}{2}$), together with the left-handed lepton doublets couple with the $\Phi_1$. The pseudo-Dirac 4-spinors $N_i$ can be written in the form of $\left[ \begin{array}{c} N_{L i}^{\text{w}} \\ i \sigma^2 N_{R_i}^{\text{w} *} \end{array} \right]$, where $N_{\text{L,R}i}^{\text{w}}$ are the sterile neutrino fields in the Weyl 2-spinor form. The corresponding Lagrangian is given by
\begin{eqnarray}
\mathcal{L}_{\text{Yukawa, Mass}}^{\nu} = -Y_{N i j} \overline{L}_{L i} \tilde{\Phi}_2 N_{R j} - m_{N i j} \overline{N}_{i} N_{j} - \mu_{i j}^1 \overline{N_{R i}^c} N_{R j} - \mu_{i j}^2 \overline{N_{L i}^c} N_{L j}, \label{PseudoDiracYukawaMass}
\end{eqnarray}
where $Y_N$ is the $3 \times 3$ Yukawa coupling constant matrix, $m_N$ is the $3 \times 3$ Dirac mass matrix between the sterile neutrino pairs, $\mu$ is a $3 \times 3$ mass matrix which softly breaks the lepton number, and $N_{L,R i}^c=-i \gamma^2 \gamma^0 \overline{N_{L,R i}^c}^T$ is the charge conjugate transformation of the $N_{L,R i}$ field.  However, as for the tree-level inverse seesaw model, there exist examples in which only the mass terms corresponding to $\mu^2_{i j} \overline{N_{L i}^c} N_{L j}$ are generated and discussed \cite{InverseTriBi1, InverseTriBi2, InverseTriBi3}. In fact, it is easier to generate the correct light neutrino mass matrix pattern in a discrete symmetry and flavon-based model if the lepton flavour violation has only one source (It, in this paper, refers to $\overline{N_{L i}^c} N_{L j}$.), though, in this paper, we discuss both the contribution from $\mu^{1,2}$ for completion.

\section{Neutrino Masses}

\begin{figure}
\includegraphics[width=3in]{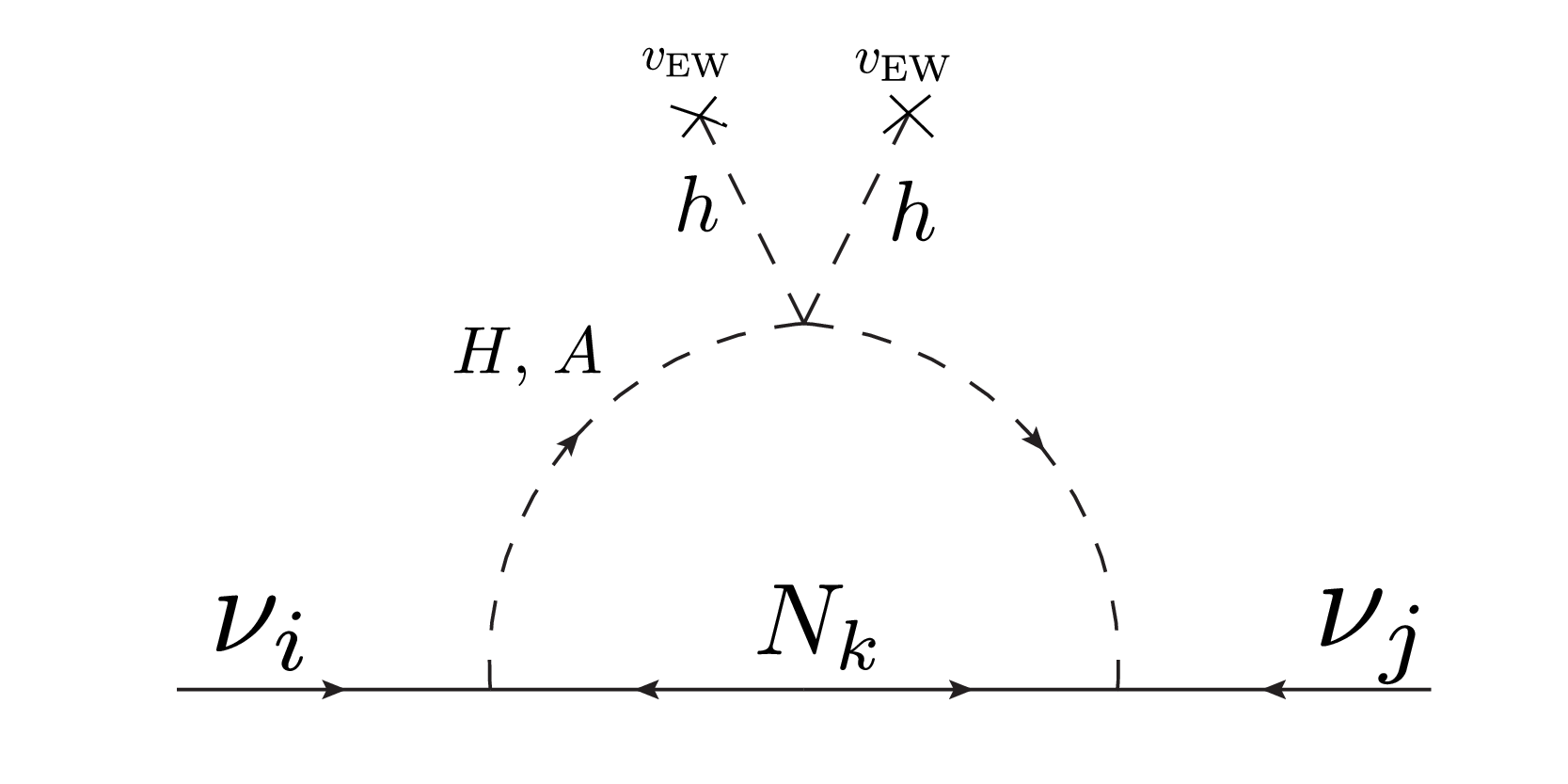}
\includegraphics[width=3in]{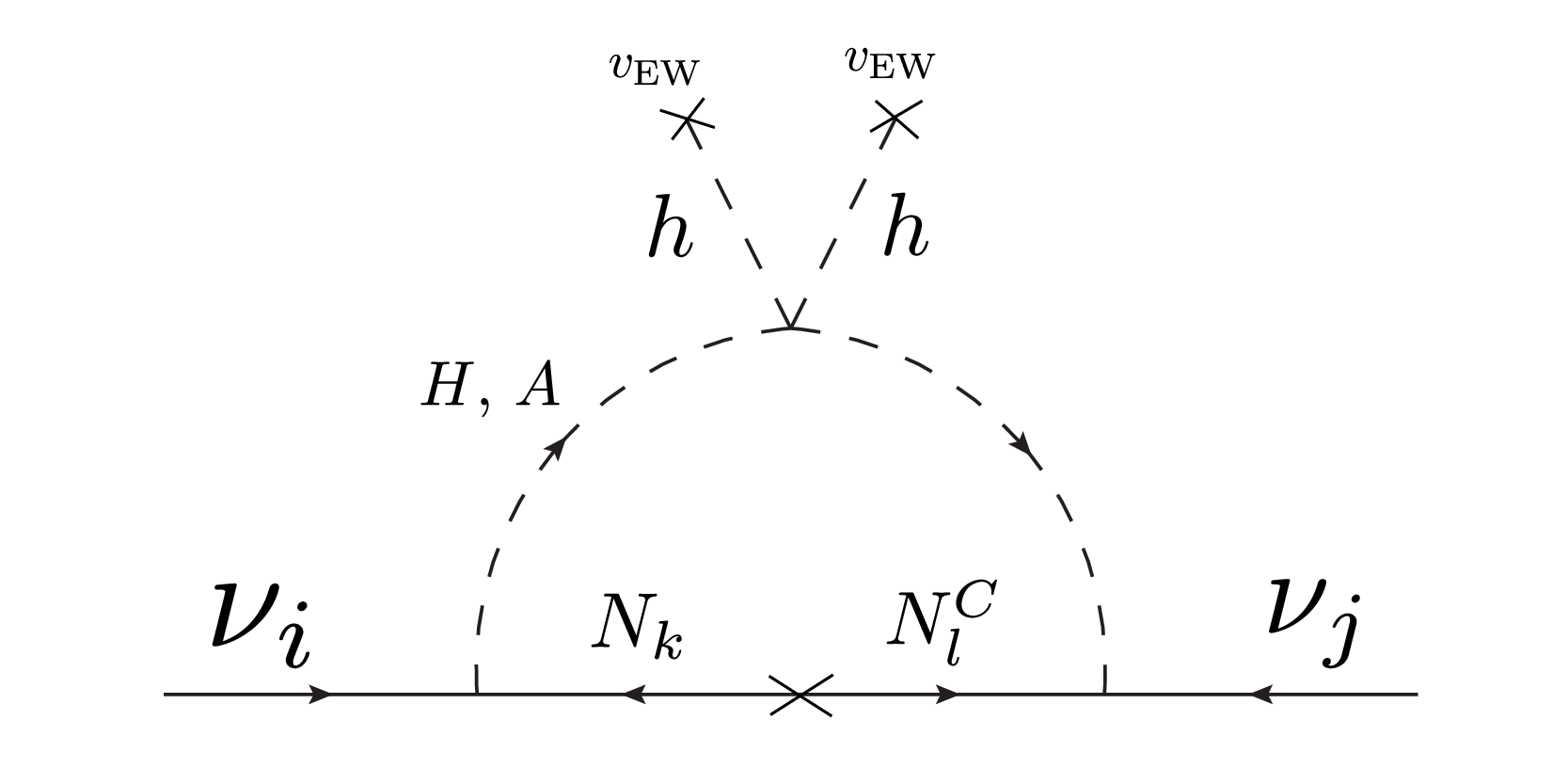}
\caption{The left panel shows the neutrino loop-induced mass in the case of Majorana sterile neutrinos. The right panel shows the case of pseudo-Dirac sterile neutrinos.}
\label{LoopMass}
\end{figure}

The right panel of the Fig.~\ref{LoopMass} shows the diagram that induces the neutrino masses. Inside the loop there is a Majonara mass insertion term originated from the Eqn.~(\ref{PseudoDiracYukawaMass}). By principle, we can directly calculate through the this diagram, however, in this paper, we adopt another method. In fact, the pseudo-Dirac neutrinos can actually be regarded as a pair of nearly-degenerate Majorana fermions. Each fermion contributes to the left panel of the Fig~\ref{LoopMass}, and by summing over all the results, a remnant proportional to the $\mu_{i j}$ is left over.

In spite of the coupling constants, the kernel of the left panel of Fig.~\ref{LoopMass} is given by the Ref.~\cite{Scotogenic1}.
\begin{eqnarray}
f(M_k, m_H, m_A) = \frac{M_k}{16 \pi^2} \left[ \frac{m_H^2}{m_H^2-M_k^2} \ln \frac{m_H^2}{M_k^2} - \frac{m_A^2}{m_A^2-M_k^2} \ln \frac{m_A^2}{M_k^2} \right], \label{IntKernal}
\end{eqnarray}
where $M_k$, $m_{H, A}$ are the mass of the Majorana sterile neutrino, and the masses of the CP-even and CP-odd neutral exotic Higgs bosons $H$ and $A$.

In the Weyl basis, the mass terms in (\ref{PseudoDiracYukawaMass}) can be written in the form of
\begin{eqnarray}
\mathcal{L}^{\text{SM}}_{\text{Mass}} &=& m_{N_{ij}} (N_{L i}^{\text{w}} N_{R j}^{\text{w}} + \text{h.c.}) +\mu^1_{i j} N_{R i}^{\text{w}} N_{R j}^{\text{w}} + \mu^2_{i j} N_{L i}^{\text{w}} N_{L j}^{\text{w}}.
\end{eqnarray}
That is to say, in the $N_{L i}$, $N_{R i}$ basis, the blocking mass matrix is given by
\begin{eqnarray}
M_N = \left[
\begin{array}{cc}
\mu^1 & m_N \\
m_N^T & \mu^2
\end{array}
\right], \label{SterileMassMatrix}
\end{eqnarray} 
where $m_N=[m_{N i j}]$ and $\mu^{1,2} = [\mu^{1,2}_{i j}]$ are the $3 \times 3$ submatrix. Without loss of generality, let $m_N$ be diagonalized with the eigenvalue $m_{Ni}$, $i=1,2,3$, and regard $\mu^{1,2}_{i j}$ as the perturbation parameter, and diagonalize (\ref{SterileMassMatrix}), the rotation matrix is given by
\begin{eqnarray}
M_N &\Rightarrow & V^T M_N V, ~V=\frac{1}{\sqrt{2}} \left[ \begin{array}{cc}
I & I \\
-I & I 
\end{array} \right] \delta, \nonumber \\
\delta &=& \left[ \begin{array}{cc}
I_n & C \\
-C^T & -I_n
\end{array} \right],
\end{eqnarray}
where
\begin{eqnarray}
I_{n i j} &=& \frac{\mu_{i j}^{-}}{2 (m_{N j}-m_{N i})} \text{ for $i \neq j$, }I_{n i j}=0 \text{ for $i=j$.} \nonumber \\
C_{i j} &=& \frac{\mu_{i j}^{+}}{2 (-m_{N j}-m_{N i})},
\end{eqnarray}
where $\mu_{i j}^{+} = \mu^{1}_{i j} + \mu^{2}_{i j}$ and $\mu_{i j}^{-} = \mu^{2}_{i j}-\mu^{1}_{i j}$. Replace each masses in (\ref{IntKernal}) with $\left( V^T M_N V \right)_{ii}$, and multiply the coupling constants $Y_{N i k} Y_{N j l}$, then sum over all the terms while drop higher orders of $\mu_{i j}$, we acquire
\begin{eqnarray}
m_{\nu i j} = \sum_{k,l=1\text{-}3} Y_{N i k} Y_{N j l} \left[ \mu_{k l}^- F(m_{N l}, m_{N k}, m_H, m_A) - \mu_{k l}^+  \frac{f(m_{N l}, m_H, m_A) + f(m_{N k}, m_H, m_A) }{m_{N l}+m_{N k}} \right], \label{NeutrinoMassSolution}
\end{eqnarray}
where
\begin{eqnarray}
F(m_{N l}, m_{N k}, m_H, m_A) = \left\lbrace
\begin{array}{cc}
\frac{f(m_{N l}, m_H, m_A) - f(m_{N k}, m_H, m_A) }{m_{N l}-m_{N k}}, & \text{ when } m_{N l} \neq m_{N k},\\
\frac{\partial f(m_{N l}, m_H, m_A)}{\partial m_{N l}}, & \text{ when } m_{N l} = m_{N k}.
\end{array} \right.
\end{eqnarray}

In this paper, we ignore all the CP phases for simplicity and adopt the central values\cite{NeutrinoGlobalFit, PDG}
\begin{eqnarray}
& & \Delta m_{21}^2 = 7.37 \text{eV}^2, ~~~~~~ |\Delta m^2| = |\Delta m_{32}^2 + \Delta \frac{m_{21}^2}{2}| = 2.50 \text{eV}^2, ~~~~~~ \sin\theta_{12}^2 = 0.297 \nonumber \\
& & \sin^2\theta_{23} = 0.437, ~~~~~~ \sin^2\theta_{13}=0.0214 \label{Neutrino_Parameters}
\end{eqnarray}
to calculate the $m_{\nu}$ through the Pontecorvo–Maki–Nakagawa–Sakata (PMNS) matrix
\begin{eqnarray}
& & U = \left[
\begin{array}{ccc}
c_{12} c_{13} & s_{12} c_{13} & s_{13} e^{-i \delta} \\
-s_{12} c_{23} - c_{12} s_{23} s_{13} e^{i \delta} & c_{12} c_{23} - s_{12} s_{23} s_{13} e^{i \delta} & s_{23} c_{13} \\
s_{12} s_{23} - c_{12} c_{23} s_{13} e^{i \delta} & -c_{12} s_{23} - s_{12} c_{23} s_{13} e^{i \delta} & c_{23} c_{13}
\end{array}\right] \times \text{diag}(1, e^{i \frac{\alpha_{21}}{2}}, e^{i \frac{\alpha_{31}}{2}}), \nonumber \\
& & \text{diag}(m_1, m_2, m_3) = U^T m_{\nu} U,
\end{eqnarray}
where $s_{ij}=\sin{\theta_{ij}}$, $c_{ij}=\cos\theta_{ij}$, and $\theta_{ij}$'s are the mixing angles. The CP-phase angle $\delta$, and the two Majorana CP phases $\alpha_{21,31}$ are omitted. $m_{1,2,3}$ are the masses of the three light neutrinos. Currently, the mass hierarchy (normal or inverse hierarchy) and the absolute neutrino masses still remain unknown. By assuming the mass hierarchy and the lightest neutrino mass $m_{\nu 0}$, matrix $m_{\nu}$ can be computed and then  $\mu$ can be calculated through inversely solving the Eqn.~(\ref{NeutrinoMassSolution}).

\section{Discussions on Oblique Parameters,  $l_1 \rightarrow l_2 \gamma$ and the Collider Constraints}

The one-loop level contributions to the Peskin-Takeuchi oblique parameters $S$, $T$, and $U$ from the general two Higgs doublet model (THDM) have been calculated in the literature \cite{THDMSTU1, THDMSTU2}. Some papers (e.g., Ref.~\cite{THDMSTUPlot1, THDMSTUPlot2} ) also plot the allowed region constrained by the oblique parameters. From the formula and the figures in the literature, we can easily find that if $m_{A} \approx m_{H^{\pm}}$, or if $m_{H} \approx m_{H^{\pm}}$, the contributions from the exotic Higgs doublets will nearly disappear in the alignment limit. Therefore, in this paper, we discuss the following benchmark parameter spaces:
\begin{enumerate}[i]
\item $m_{H}=m_{H^{\pm}}$, $m_{A} \leq m_{H^{\pm}}$,
\item $m_{A}=m_{H^{\pm}}$, $m_{H} \leq m_{H^{\pm}}$.
\end{enumerate}
Although $m_{H,A}>m_{H^{\pm}}$ is also possible, however, this will make some parameters decouple and we aim at discussing as much phenomenology (allowed by the current constraints) as possible in this paper. In this paper,we do not discuss such parameter space here.

\begin{figure}
\includegraphics[width=3in]{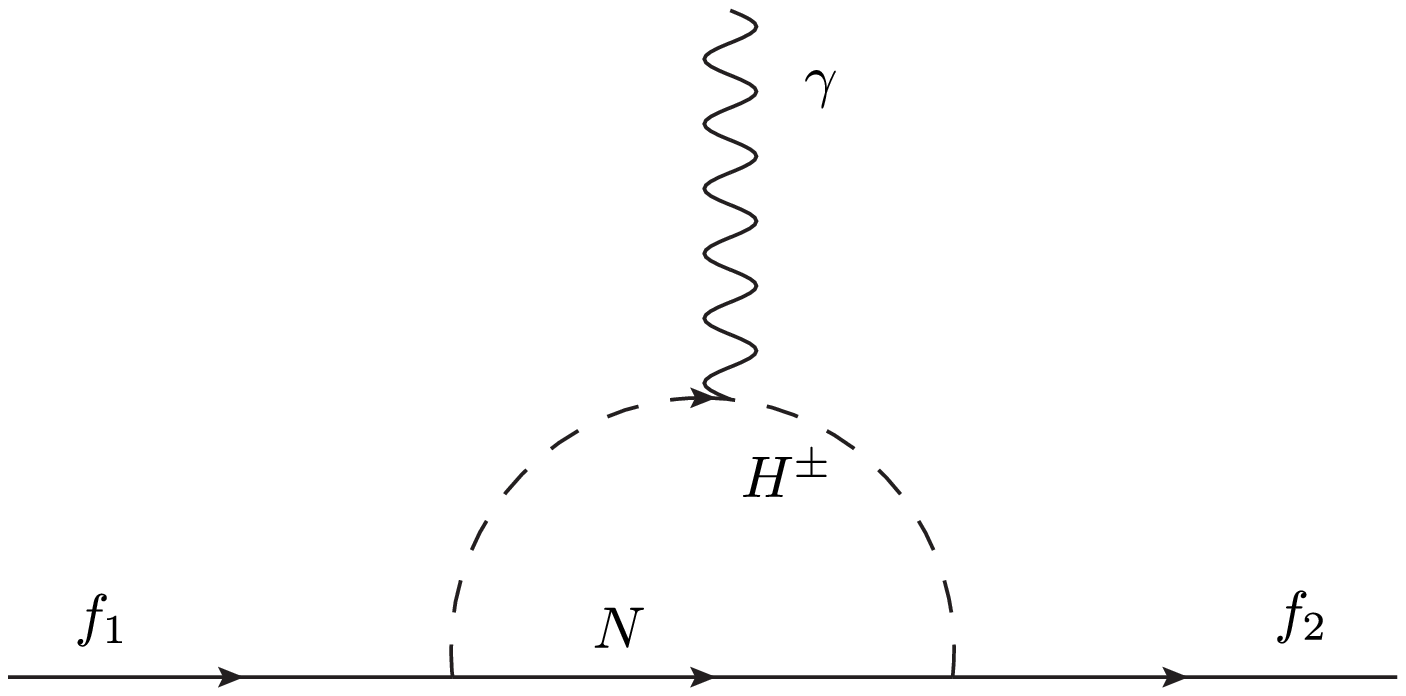}
\includegraphics[width=3in]{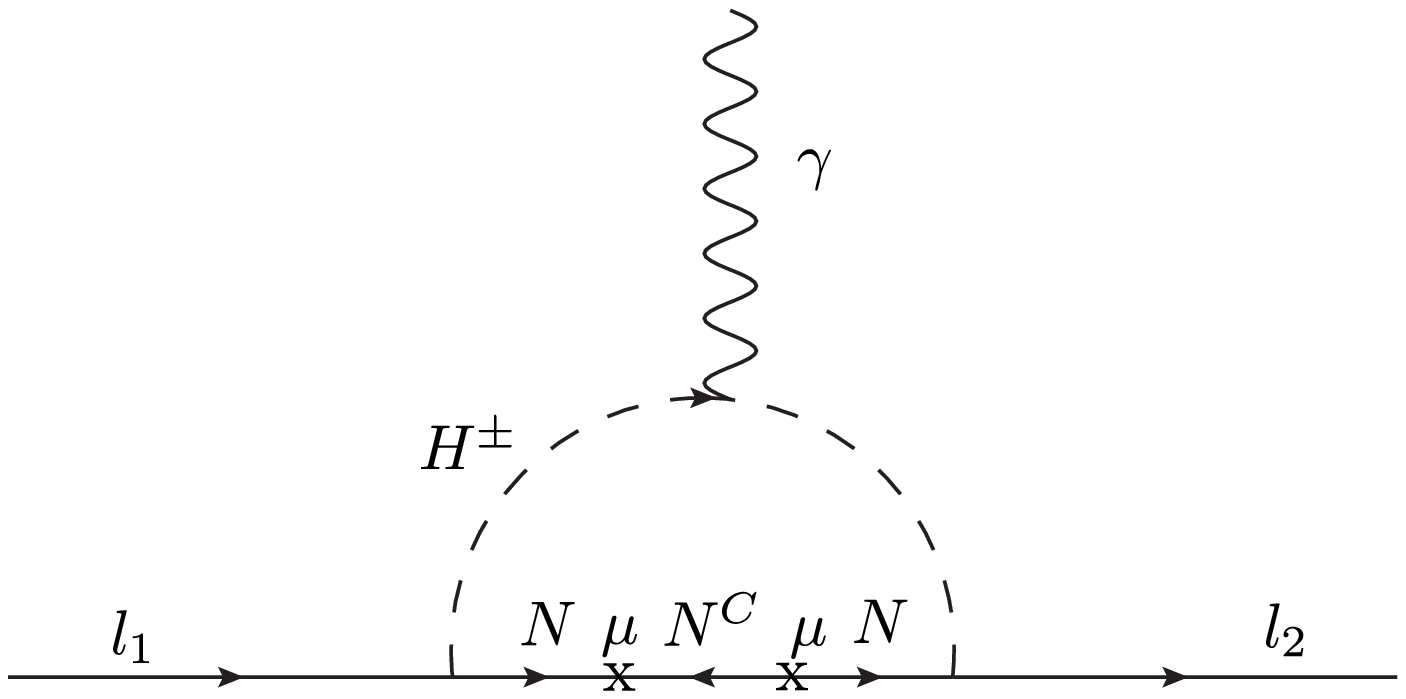}
\caption{The $l_1 \rightarrow l_2 \gamma$ diagrams.} \label{l1l2gamma}
\end{figure}

The leptonic flavour changing neutral current (FCNC) decays $l_1 \rightarrow l_2 + \gamma$ in Fig.~\ref{l1l2gamma} set constraints on the parameter space. In the original scotogenic model, Ref.~\cite{ScotogenicRequired} had pointed out that in the usual scotogenic model, the parameter space is quite constrained by the $\mu \rightarrow e \gamma$ bounds. The nearly degenerate neutrino mass scenario to avoid this bound has become very unfavorable considering the recent cosmological bound on the neutrino masses \cite{Planck} together with the oscillation data. Similar to the cases in the Ref.~\cite{MyPaper}, the FCNC elements in the $Y_{N i j}$ and the $m_{N i j}$ are stringently bounded through the diagram in the left panel of the Fig.~\ref{l1l2gamma}.  Setting $m_N \propto I$ and $Y_{N i j} \propto I$ will simply avoid this problem, where $I$ is the identity matrix. This is not ad-hoc, if some flavon-based inverse-seesaw models like Ref.~\cite{InverseTriBi1, InverseTriBi2, InverseTriBi3} that can describe the origin of these parameters are transplanted to our loop-level case. Eventually, if all of the leptonic FCNC effects originates from the $\mu$ terms, the diagram in the right panel of Fig.~\ref{l1l2gamma} will become the lowest order of contributions to the $l_1 \rightarrow l_2 + \gamma$ and become severely suppressed by the factor $\sim \frac{\mu^4}{m_N^4}$. Therefore, in this paper, we only consider the case that $m_N \propto I$, $Y_{N i j} \propto I$, $\mu \propto \!\!\!\!\!/ ~I$.

In this paper, we are also interested in the case that the fermionic $N$'s are the lightest $Z_2$-odd particles that turn out to be the candidate of the dark matter. $H$, $A$ are not considered partly because such cases have been widely and sufficiently talked about in the literature. 

\begin{figure}
\includegraphics[width=5in]{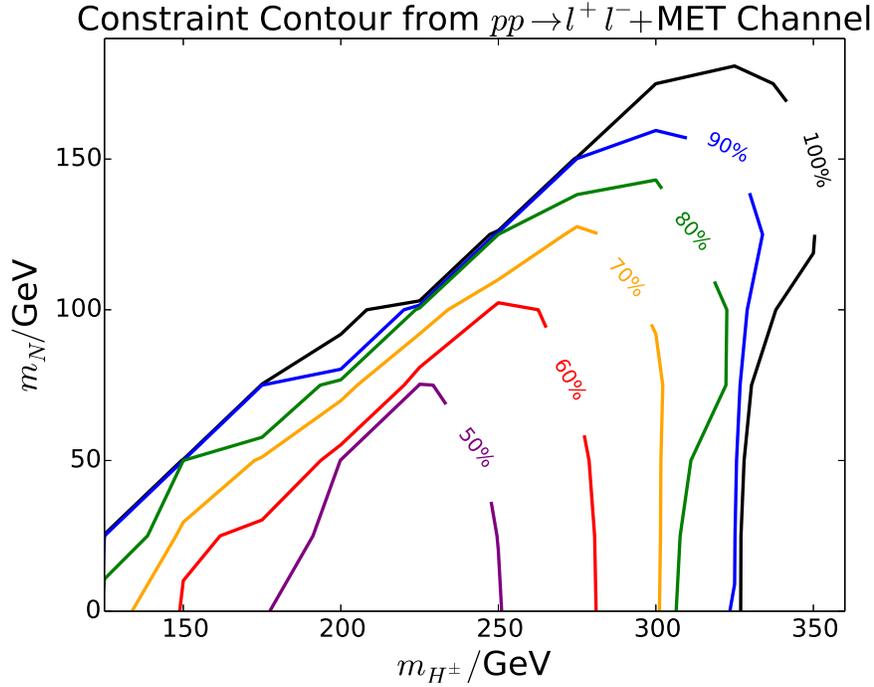}
\caption{95\% CL.~exclusion limits on the $m_{H^{\pm}}$-$m_{N}$ parameter space. The Results on different branching ratios of $H^{\pm} \rightarrow N_i l_j^{\pm}$ (printed on each curve in the panel) are plotted.} \label{Constraint_Altogether}
\end{figure}

On the collider, usually $H^{\pm}$, $H$ and $A$ are produced by the electro-weak processes and then decay into the missing energy plus some SM final states. Notice that the decay rate between different sterile neutrino $N_i$'s are so severely suppress by the smallness of their mass differences due to $m_N \propto I$ and $\mu \ll m_N$, that we assume such decay will never happen inside the detector.  In fact, even if they can decay inside the detector, this will only produce some rather soft objects that might be difficult to figure out. Here we regard all the $N_i$'s as the missing energies (ME), and we have examined various combinations of the production processes $p p \rightarrow H^+ H^-$, $p p \rightarrow H^{\pm} H$, $p p \rightarrow H^{\pm} A$, $p p \rightarrow H A$ with the various decay channels $H^{\pm} \rightarrow N_i l_j^{\pm}$, $H^{\pm} \rightarrow A/H+W^{\pm}$, $H(A) \rightarrow A(H) +Z$, etc. The LHC experiments have extracted the data on some of the channels, and among these we select the most stringent one, which is $p p \rightarrow H^+ H^- \rightarrow l^+ l^- +$ME.  For this final state, the ATLAS collaborator have provided the binned data of the SM background in the Ref.~\cite{ATLAS_SleptonResult}. Here, we implement the model files and generate the events by FeynRules 2.3.28\cite{FeynRules}+MadGraph 2.5.5\cite{MadGraph}, and bin our results by Madanalysis5.1.5 \cite{MA1, MA2, MA3}. Both the same flavour (SF) and the different flavour (DF) data are considered. Since the final states are leptons, only parton-level analyses are processed. We have calculated the CL.~ ratio according to the Ref.~\cite{CLCalc}, and scanned in the $m_{H^{\pm}}$-$m_{N}$ parameter space. We plot our results of 95\% CL.~exclusion limits in the Fig.~\ref{Constraint_Altogether}.

\section{Calculations of some Observables}

In this section, we aim at calculating the following observables:
\begin{itemize}
\item Relic density of the dark matter.
\item Shiftings on the Z-resonance observables $R_l = \frac{\Gamma_{Z \rightarrow \text{hadrons}}}{\Gamma_{Z \rightarrow l^{+} l^{-}}}$, $A_l = \frac{2 \overline{g}_V^l \overline{g}_A^l}{\overline{g}_V^{l 2}+\overline{g}_A^{l 2}}$.
\item Shiftings on the $Z \rightarrow $invisible parameter $N_{\nu}^l = \frac{\Gamma_{\text{inv}}^Z}{\Gamma_l^Z} \left( \frac{\Gamma_l^Z}{\Gamma_{\nu}^Z} \right)_{\text{SM}}$.
\end{itemize}

The review on the $R_l$, $A_l$, and $N_{\nu}^l$ can be found in Ref.~\cite{PDG}. The relic density of the dark matter is calculated by micrOMEGAs 4.3.5 \cite{MicrOMEGAs1, MicrOMEGAs2}, with the our model file exported by FeynRules 2.3.28. The inert Two Higgs Doublet Model part of the model file is based on the Ref.~\cite{Inert1, Inert2}.

The shiftings on the electroweak parameters $\delta R_l$, $\delta A_l$, and $\delta N_{\nu}^l$ are calculated according to the formulas and steps listed in Ref.~\cite{MyPaper}, where the one-loop corrections to the $Z$-$l$-$l$ coupling constants are computed and then replace their values to the expressions of $\delta R_l$, $\delta A_l$, and $\delta N_{\nu}^l$ which depend on them. The computing processes can be compared and checked with the Ref.~\cite{ZbbCalc}. Note that in the case of this paper, neither the tree-level correction to the muon's decay constant $G_F$ nor the tree-level mixings between the sterile neutrinos and the light neutrinos  exists, therefore the computing procedures are much simpler than those in the Ref.~\cite{MyPaper}.

\begin{figure}
\includegraphics[width=5in]{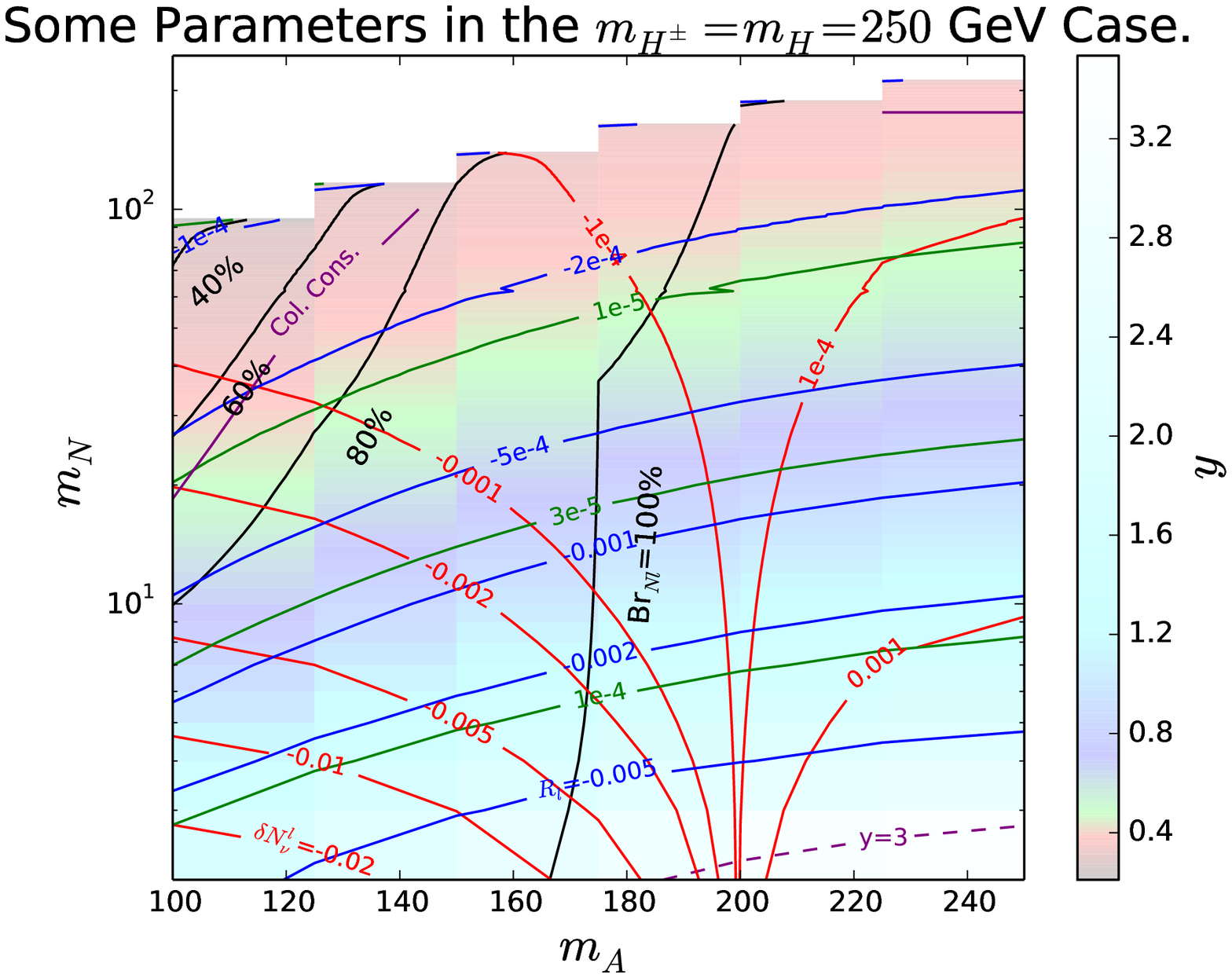}
\includegraphics[width=5in]{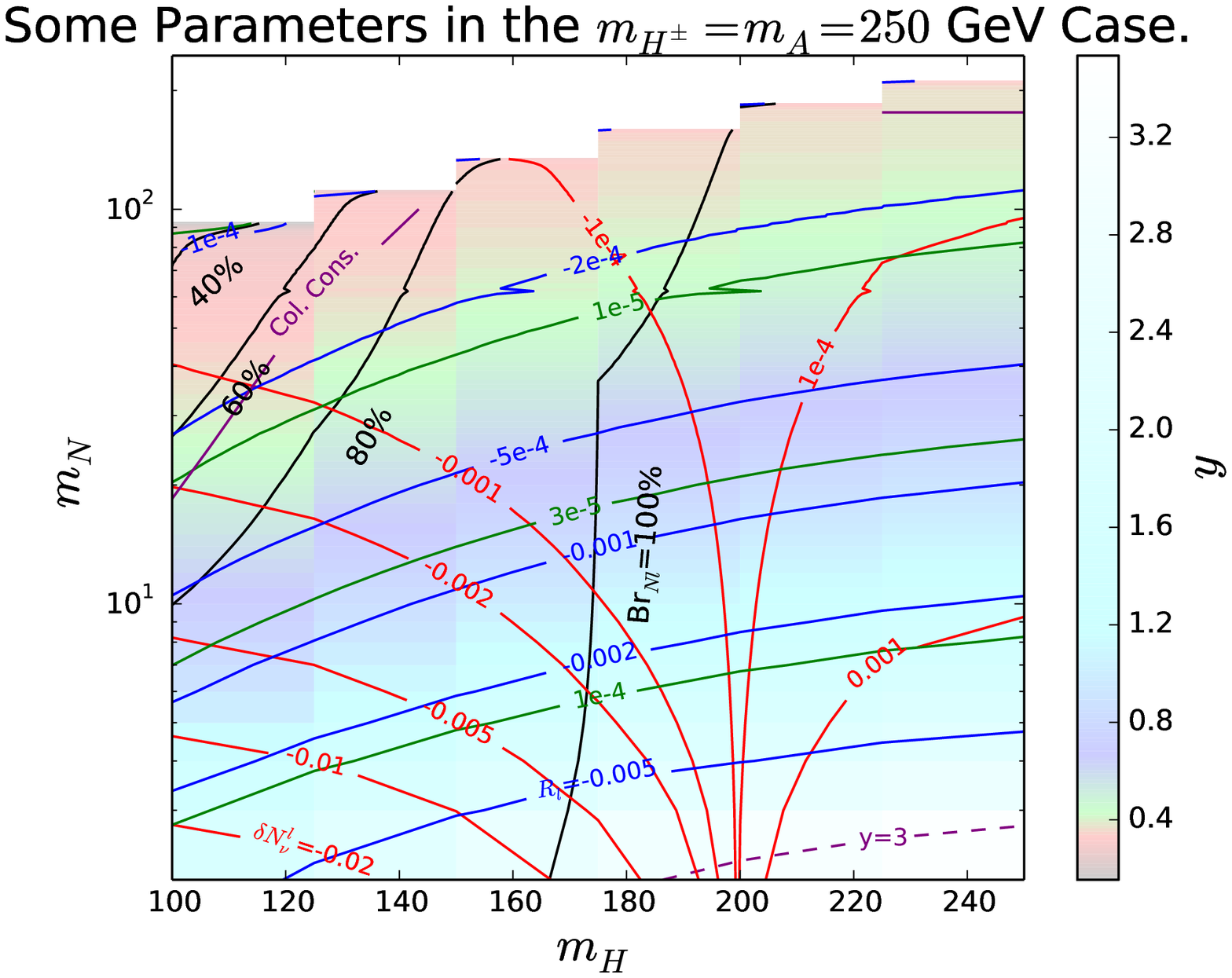}
\caption{$\delta R_l$ (blue line), $\delta A_l$ (green), $\delta N_{\nu}^l$ (red) in the $m_{H^{\pm}} = 250$ GeV case. The color in the background indicate the Yukawa coupling $y$ that can result in the correct relic density.  The dashed purple line in the right-bottom of the picture shows the $y = 3 \sim \sqrt{4 \pi}$ boundary of the Yukawa coupling. The black lines indicate the branching ration of the $ H^{\pm} \rightarrow N_{i} l^{\pm}_j$ decay channel, and the solid purple line marked with ``Col. Cons.'' gives the CL.~95\% collider constraint derived from the Fig.~\ref{Constraint_Altogether}. The right-down direction from this line has been excluded.}
\label{Result_250}
\end{figure}

\begin{figure}
\includegraphics[width=5in]{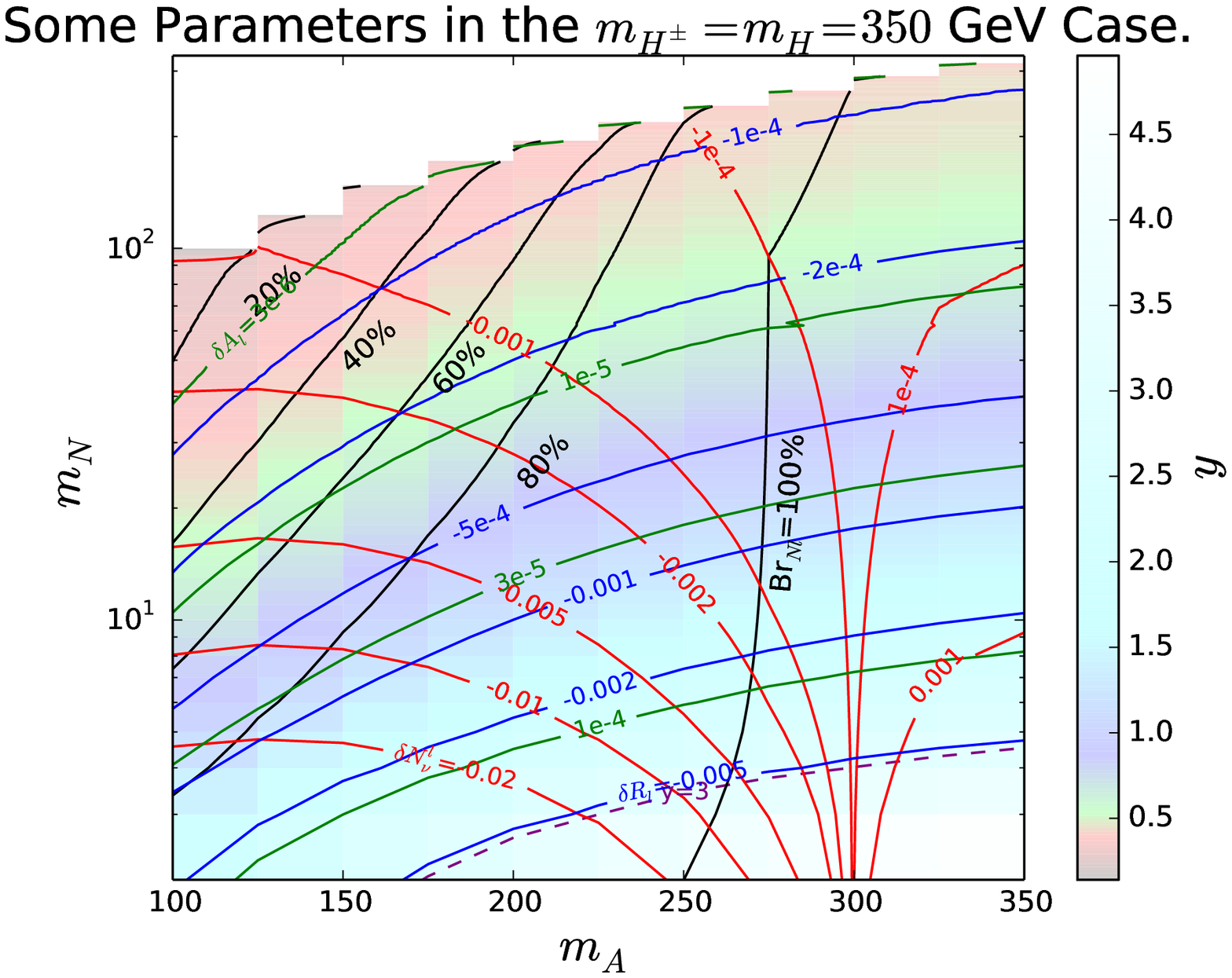}
\includegraphics[width=5in]{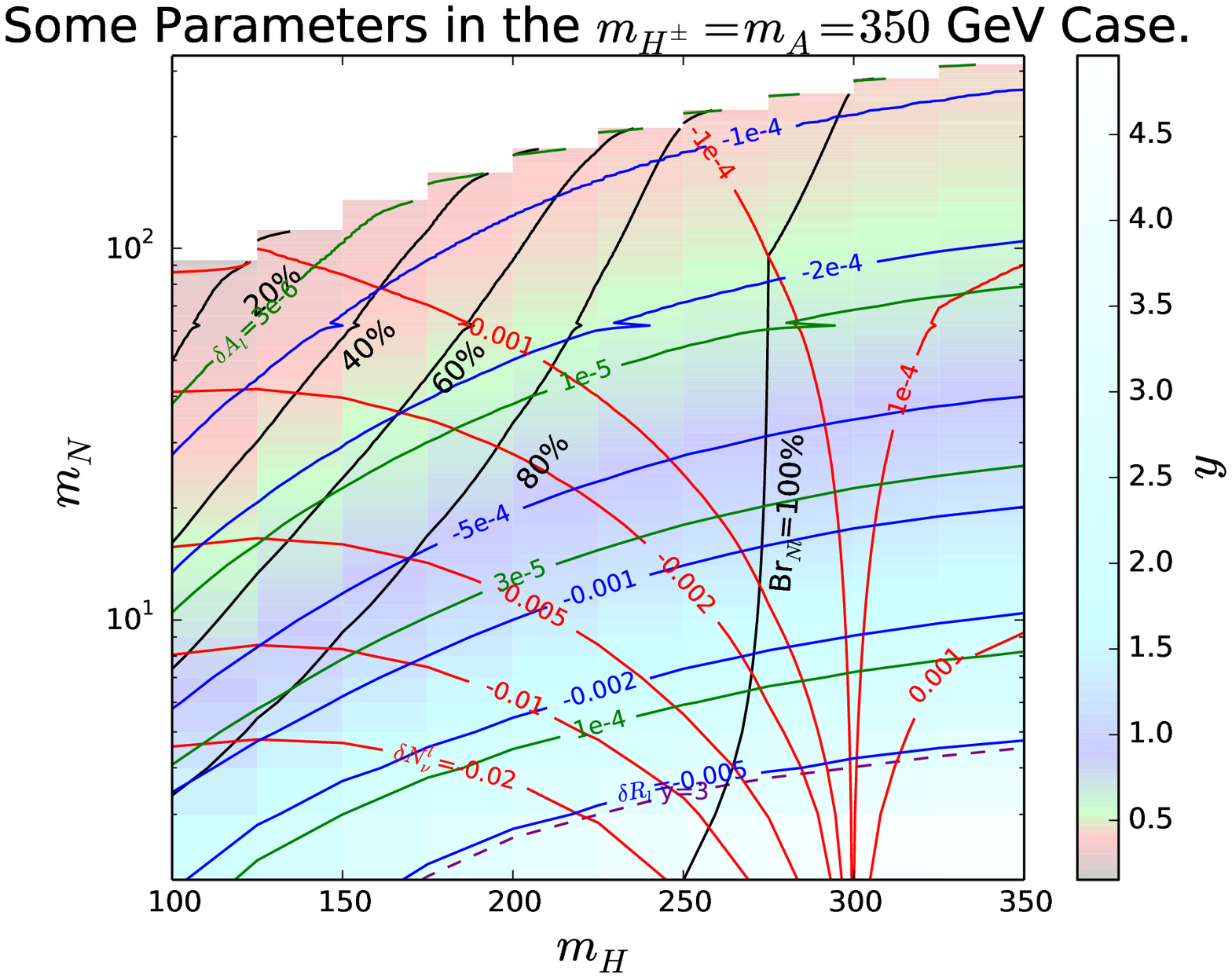}
\caption{Results in the $m_{H^{\pm}} = 350$ GeV case. The meanings of the colors and the lines are similar to the Fig.~\ref{Result_250}.}
\label{Result_350}
\end{figure}

\begin{figure}
\includegraphics[width=5in]{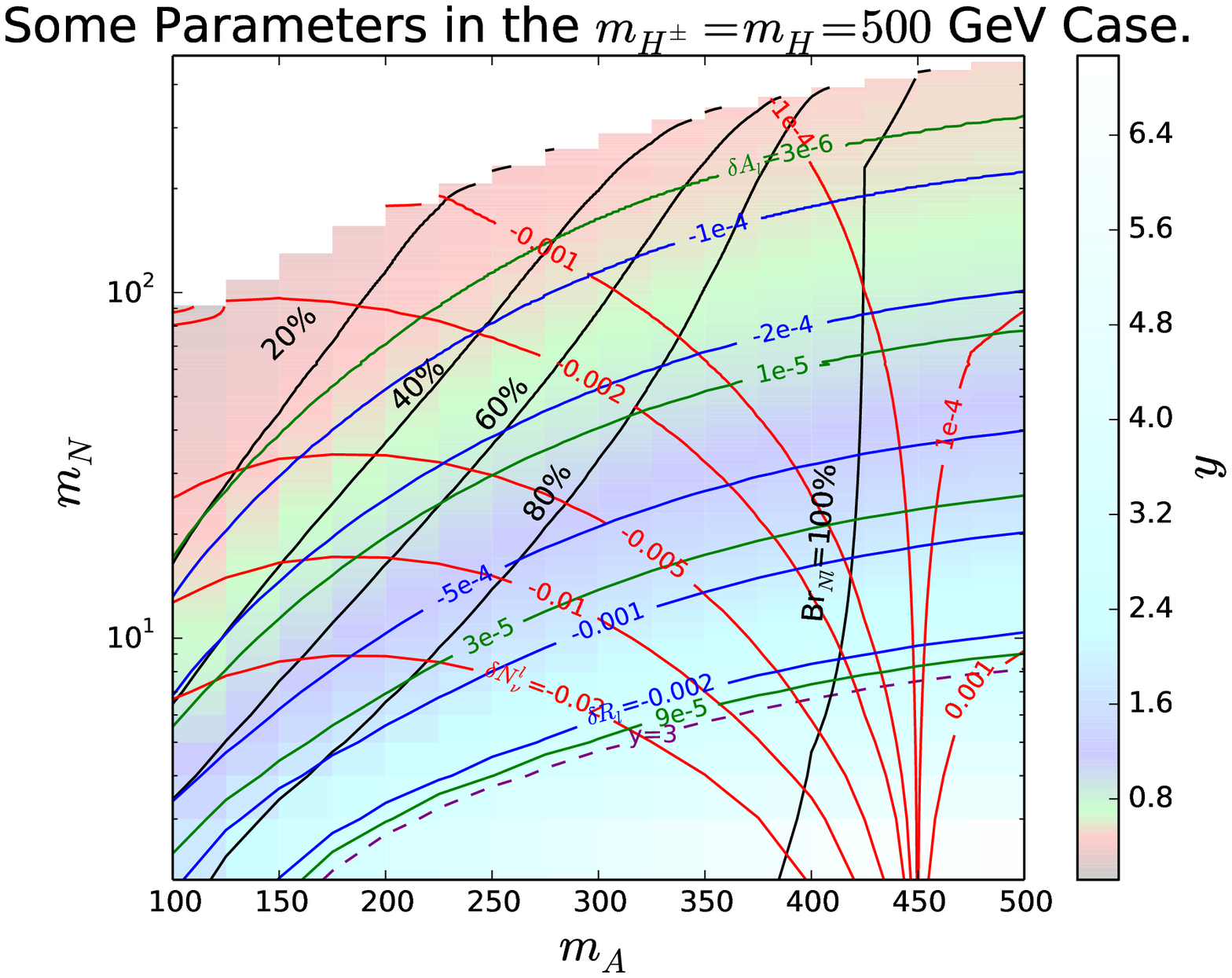}
\includegraphics[width=5in]{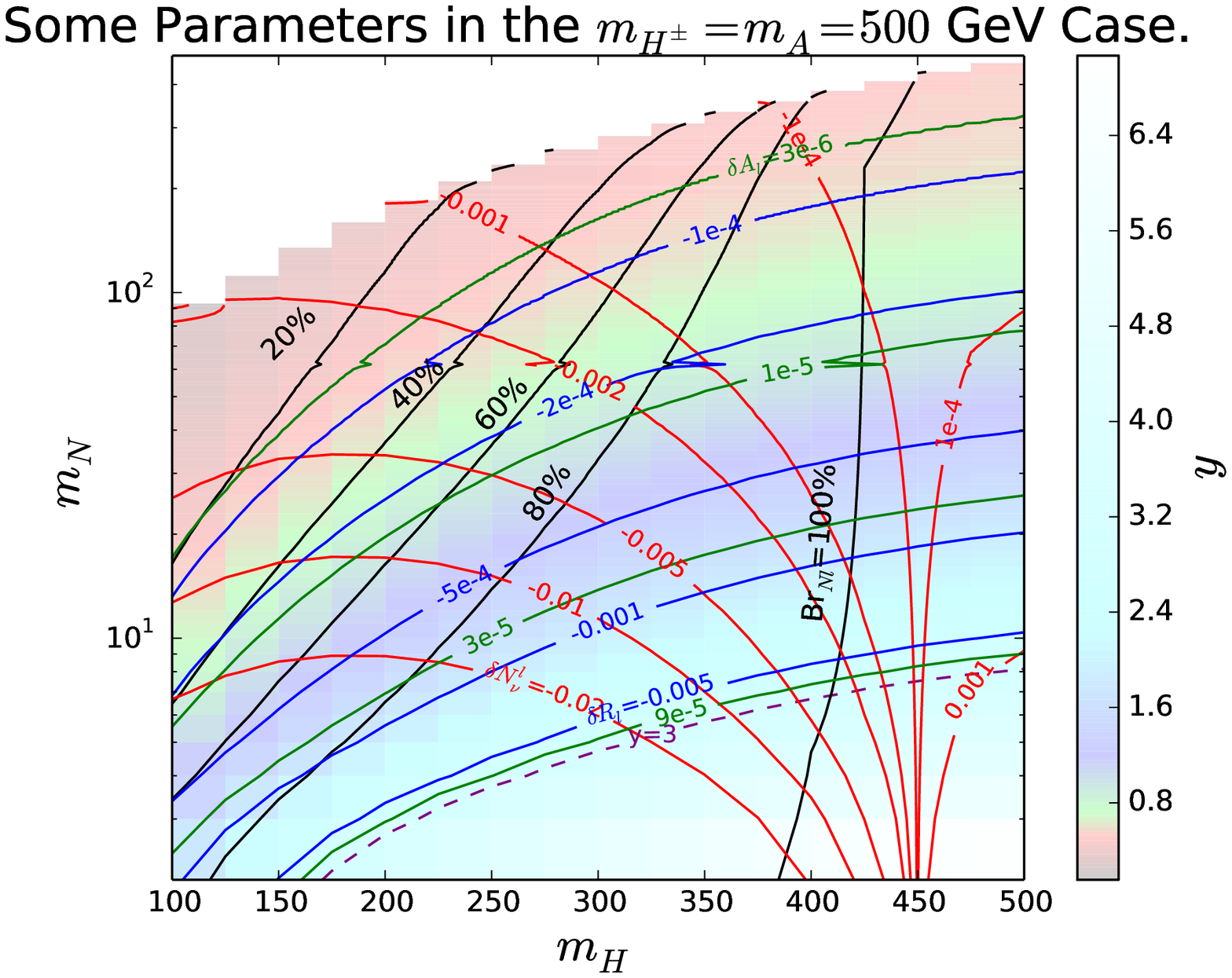}
\caption{Results in the $m_{H^{\pm}} = 500$ GeV case. The meanings of the colors and the lines are similar to the Fig.~\ref{Result_250}.}
\label{Result_500}
\end{figure}

In order to present our results, we only consider the sub parameter space of $m_{H^{\pm}}=250$, $350$, $500$ GeV.  The Yukawa coupling constant $[Y_{N i j}] = y I$ is adjusted in order for the relic density $\Omega_{\text{DM}} h^2$ to approach $0.1199 \pm 0.0027$ \cite{PlanckRelic}. Combined with the two cases in the last section, we show six plots in the Fig.~\ref{Result_250}, \ref{Result_350} and \ref{Result_500}. 


As has been mentioned, we are only interested in the case when $m_N < m_{H, A}$, therefore the upper-left part of the plots are all left blank. The color boundary  becomes a step-like shape due to the insufficient density of points on the horizontal axis and our limit on the computational resources.

\section{Discussions}

There are plans on the future experiments to measure $R_l$, $A_l$, $N_{\nu}^l$ \cite{CEPC_PreCDR, ILC_TDR_2, ZFactory, FCCee}. Currently, the collider experiments have imposed very stringent bounds on the mass of $m_H^{\pm}$. From Fig.~\ref{Constraint_Altogether}, we can easily see that $m_H^{\pm} \lesssim 325$ GeV has been excluded in the case when $m_N \rightarrow 0$ and Br$_{H^{\pm} \rightarrow N l^{\pm}}=$100\%. When Br$_{H^{\pm} \rightarrow N l^{\pm}}<$100\%, bounds on $m_{H^{\pm}}$ can be somehow relaxed. However, besides the leptonic channel, $H^{\pm}$ can only decay into $H/A+W^{\pm}$, which requires $m_{H^{\pm}}-m_{H/A} \gtrsim 100 \text{ GeV}$.  This substantially compresses the unconstrained parameter space in the case when $m_H^{\pm} \lesssim 325$. As an example, it is obvious that in the Fig.~\ref{Result_250}, only a fraction in the upper-left part of the plot remains unconstrained.

As for the $\delta R_l$, $\delta A_l$, all the corresponding diagrams only involve $H^{\pm}$ , Thus they are highly suppressed due to the constraints on $m_{H^{\pm}}$. As has been discussed in the Ref.~\cite{MyPaper}, taking both the experimental and theoretical uncertainties into account, $|\delta R_l|$ should be $\gtrsim 0.001$ and $|\delta A_l|$ should be $\gtrsim 8 \times 10^{-5}$ in order to for the new physics effects to be observed on the future experiments that measure the electroweak parameters. When $m_{H^{\pm}} > 325$ GeV, for example as in the Fig.~\ref{Result_350} and \ref{Result_500}, in some cases the predicted $|\delta R_l|$ or $|\delta A_l|$ might reluctantly reach this bound, however in this case the Yukawa coupling $y$ approaches the perturbative constraint $y<\sqrt{4 \pi} \sim 3$. On the other hand, when $m_{H^{\pm}} \lesssim 350$ GeV, for example when $m_{H^{\pm}}=250 $ GeV as shown in Fig.~\ref{Result_250}, the unconstrained parameter space usually refers to a too-small Yukawa coupling for the enough $|\delta R_l|$ or $|\delta A_l|$. Therefore, it is difficult to detect the effects on $\delta R_l$  and $\delta A_l$ from this model on future experiments.

However, $\delta N_{\nu}^l$ can be large if $m_H$ or $m_A$ are relatively small. Ref.~\cite{FCCee} has shown us that $\delta N_{\nu}^l$ can reach a statistical uncertainty of 0.00004 and a systematic uncertainty of 0.004 in its Table 1. However, The discussions in the section 7 mentioned that a desirable goal would be to reduce this uncertainty down to 0.001. In this case, the FCC-ee is enough to cover much of the parameter space in $m_{H^{\pm}} =350$, $500$ GeV. However, as has been shown in Fig.~\ref{Result_250}, it is still difficult to detect $\delta N_{\nu}^l$ in its unconstrained parameter space when $m_{H^{\pm}}=250$ GeV. Interestingly, current LEP results $N_{\nu}^l = 2.984 \pm 0.008$ show a 2-$\sigma$ deviation from the standard model prediction. If this will be confirmed in the future collider experiments, it will become a circumstantial evidence to this model.

Finally, we should discuss about the direct detection on this model. Since we have only talked about the case of the sterile neutrino playing the role of dark matter, there is no tree-level diagram for the nucleon-dark matter interactions. Ref.~\cite{CaoDirectDetection} has calculated the loop-induced $Z$-portal cross sections for the nucleon-dark matter collisions in both the Majorana and the Dirac case. In our paper, the dark matter particle is the lightest Majorana mass eigenstate of a group of pseudo-Dirac fermions. The mass difference $\delta$ between the two Majorana elements of one pseudo-Dirac pair of fermions is controlled by $\mu$. According to (\ref{NeutrinoMassSolution}), $\mu$ is inversely proportional to $F(m_{N l}, m_{N k}, m_H, m_A) - \frac{f(m_{N_l}, m_H, m_A)+f(m_{N_k}, m_H, m_A)}{m_{N_l}+m_{N_k}}$, and $f(m_{N_{l, k}}, m_H, m_A)$,  $F(m_{N_l}, m_{N_k}, m_H, m_A)$ approaches to zero in the limit $m_N \rightarrow 0$ or $m_A \rightarrow m_H$. Generally speaking, $\mu \sim 100$ eV if no particular pattern of values are appointed to $m_{H, A, N}$. For example, if the lightest left-handed neutrino is $m_{\nu 0}=0.03$ eV in the case of the normal mass hierarchy, and assign $m_H=350$ GeV, $m_A=200$ GeV, $m_N=100$ GeV, $y=0.447$, therefore $\mu_{i i} \simeq 100$ eV and $\mu_{i j} \simeq 10$ eV $(i \neq j)$. However, if $m_A \simeq m_H$, $\mu$ will be substantially amplified. For example, when $m_H=350$ GeV, $m_A=349.5$ GeV, $m_N=100$ GeV, $y=0.605$, in this case $\mu \simeq 10$ keV. Furthermore, if we again appoint $m_N=10$ GeV, and let $y=1.88$ in order for a correct relic density, $\mu$ will become $\simeq 100$ keV. Therefore, In this model, the mass splitting $\delta$ can be large enough ($\gtrsim 100$ keV) so that the dark matter can be regarded as a pure Majorana particle in some parameter space, while in some parameter space, the mass difference $\delta$ might become so small ($\sim 100$ eV) so that the dark matter can transfer between the two mass eigenstates during the collision with the nucleons. The latter case is more similar to the Dirac case discussed in the Ref.~\cite{CaoDirectDetection}. According to \cite{CaoDirectDetection}, The Z-portal spin-dependent cross section was calculated to be less than $10^{-41} \text{ cm}^2$ in both the Dirac case and the Majorana case,  while the spin-independent cross section in the Dirac case was calculated to be $\lesssim 10^{-47} \text{ cm}^2$ when $m_N < 200$ GeV. Although these are still below the experimental bounds \cite{PandaX, LUXSD, LUX, XENON1T}, it is hopeful for the future experiments to cover some of our parameter space since the current bounds are not far from the predictions.

\section{Conclusions}

We have discussed some phenomenologies of a simple inverse seesaw scotogenic model by calculating the electroweak parameters $R_l$, $A_l$. $N_{\nu}^l$ in the case of a correct dark matter relic density. The current ATLAS results have imposed stringent bounds on the parameter space, lowering the predicted $R_l$ and $A_l$. Considering both the experimental and theoretical uncertainties, it is difficult to detect the effect from this model on  $R_l$, $A_l$ in the future measurements. However, $\delta N_{\nu}^l$ can become large enough, shedding lights on verifying or constrain this model in the future.

\begin{acknowledgements}

We thank for Ran Ding, Pyungwon Ko, Peiwen Wu for helpful discussions. This work was supported by the Korea Research Fellowship Program through the National Research Foundation of Korea (NRF) funded by the Ministry of Science and ICT (2017H1D3A1A01014127), and is also supported in part by National Research Foundation of Korea (NRF) Research Grant NRF-2015R1A2A1A05001869. We also thank the Korea Institute for Advanced Study for providing computing resources (KIAS Center for Advanced Computation Abacus System) for this work.

\end{acknowledgements}

\newpage
\bibliography{InversevTHDM}
\end{document}